\documentclass{aps2010} 

\usepackage{times}
\usepackage{rawfonts}
\usepackage{graphicx}

\title{Element gain drifts as an imaging dynamic range limitation\\in PAF-based
interferometers}

\author[org1]{\textbf{\underline{\emph{O.\ M.\ Smirnov}}}}
\author[org2]{\textbf{\emph{M.\ V.\ Ivashina}}}

\address[org1]{ASTRON, P.O. Box 2, Dwingeloo, 7900AA, The Netherlands, smirnov@astron.nl}
\address[org2]{Dept. of Earth and Space Sciences, Chalmers University of Technology, S-41296 Gothenburg, Sweden, marianna.ivashina@chalmers.se}

\begin{document}

\maketitleblock


\section{Introduction}

\begin{figure}
\begin{tabular}{@{}ccc@{}}
\includegraphics[width=5cm]{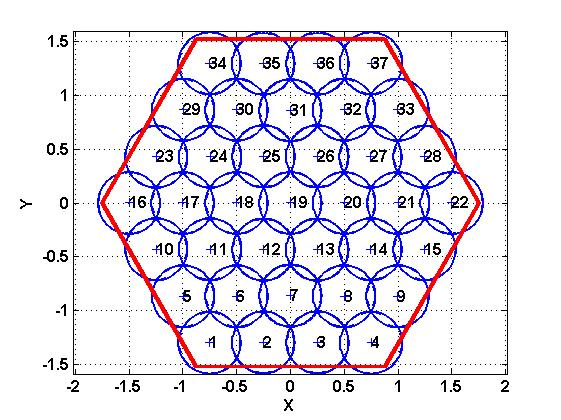} &
\includegraphics[width=5cm]{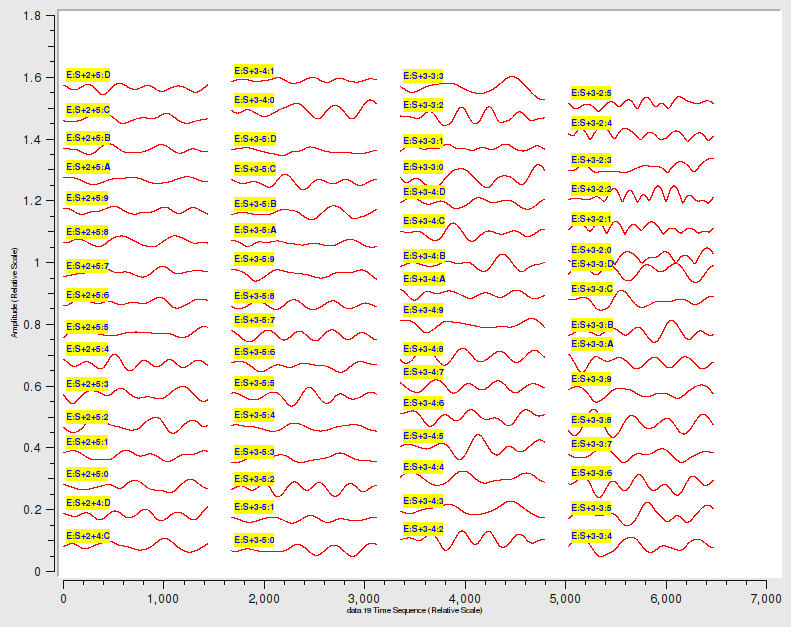} &
\includegraphics[width=5cm]{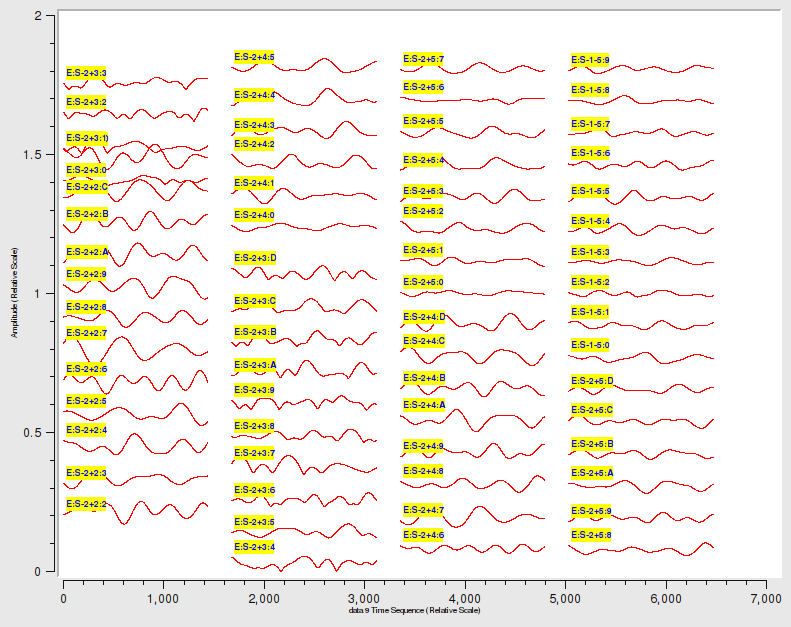} \\
(a)&(b)&(c)
\end{tabular}
\caption{\label{fig:gains}(a) Distribution of the 37 compound beams over the full field-of-view; (b,c) An example of the compound beam gains produced during the simulation. Each track shows the gain-amplitude in the direction of one source (direction), per one WSRT station, as a function of time. Note the significantly different behaviour between sources and stations.}
\end{figure}

\noindent APERTIF (“APERture Tile In Focus”) [1] is a novel technology front end system that is currently being developed by ASTRON with the aim of replacing the conventional single-pixel horn feeds of the Westerbork Synthesis Radio Telescope (WSRT) by Phased Array Feeds (PAFs). These PAFs are comprised of 100+ small antenna elements and can form multiple beams on the sky for each antenna pointing direction, and hence significantly improve the survey speed of the current instrument. The design calls for 37 closely overlapping beams per each antenna (Fig.~\ref{fig:gains}a), with an effective field of view of 8 square degrees. At present, a prototype called DIGESTIF, consisting of a PAF with 121 Vivaldi antennas connected to low noise amplifiers (LNAs) and a digital beamforming network, has been installed at one of the WSRT's 14 reflector antennas, and is being tested. This system operates at frequencies ranging from 1 to 1.75 GHz, with an instantaneous bandwidth of 300 MHz. The beamforming is performed off-line based on the measured correlations between the receiver channels [2]. Recently, DIGESTIF was equipped with an auxiliary calibrator system that can be used to track electronic gain variations of the PAF receiver during the observations [3]. Random drifts in these gains cause temporal instability of the compound beam patterns, which in turn affects interferometric imaging performance. In this paper, we describe a simulation framework for exploring this issue, and exercise it to determine whether these drifts can constitute an imaging dynamic range limitation for APERTIF.

\section{Beamforming and element gain drifts}

\noindent An electromagnetic field (EMF) propagating as a transverse plane wave, arriving from any given direction on the sky $\vec\sigma$, can be 
instantaneously characterized by a two-vector of complex amplitudes $\vec \epsilon = (\epsilon_x,\epsilon_y)^T$. The nominal ``voltage'' response $v$ of a single compound beam (composed from $n$ PAF elements) for the given direction can then be described by a chain of matrix products: $v=\vec w^T \mathbf{G} \mathbf{E} \vec\epsilon$, where $\mathbf{E}$ is a $n\times2$ matrix of direction-dependent beam gains associated with each PAF element, $\mathbf{G}$ is a diagonal matrix of LNA gains (one per element), and $\vec w$ is a vector of beamformer weights. A beamformer can simultaneously apply different weight vectors $\vec w$ to produce compound beams that point in different directions on the sky. The 37 synthesized beams actually require a total of 74 compound beams, since polarization measurements require a pair of orthogonally polarized beams (preferably equally sensitive to the corresponding orthogonal components of the incident EM field) per each direction. In an interferometer, the signals from each pair of beams are then correlated between every station of the interferometer array, forming up a total of four complex visibilities per baseline ($xx$, $yy$, $xy$ and $yx$), per 37 directions. These visibilities can be described using the Jones or Mueller formalism of the radio interferometry measurement equation (RIME) [4].

Beamformer weights may be chosen to optimize different aspects of the resulting compound beams, such as sensitivity, SNR, shape, instrumental polarization, etc. This area is undergoing intensive development, and there is as yet no clear consensus of what constitutes an optimal beam. For the purposes of this study, we concern ourselves with a different question: given a beamforming strategy, how does the behaviour of $\mathbf{G}$ influence our compound beam, and how (or if) this in turn affects our ability to reconstruct images from the measured visibilities. For purposes of illustration, we will adopt a beamforming strategy that seeks maximize the SNR of the compound beams [2], but the same approach may be applied to evaluate any other strategy.

The LNA gains given by $\mathbf{G}$ are in principle unknown, and drift in time (independently of each other) in response to changing environmental conditions, resulting in what are called element gain drifts (EGDs). During initial beamformer calibration, the current element gains are estimated, and a suitable weight vector is chosen to compensate (so that the product $\vec w^T\mathbf{G}$ is equivalent to a set of ``ideal'' weights that would be used if the gains were constant). From that point, $\mathbf{G}$ begins drifting, resulting in a compound beam that slowly distorts in an difficult-to-predict manner (Fig.~\ref{fig:gains}b). Classic interferometry (and the self-calibration algorithm in particular) implicitly assumes that all stations of an interferometer array have an identical and stable beam response. EGDs violate this assumption, resulting in a time-variable, station-dependent, direction-dependent effect (DDE). This in turn produces subtle (or not so subtle) artefacts in interferometric images; see [5] for a more detailed look at this error mechanism. If these artefacts are sufficiently high (compared to the thermal noise), they could effectively limit the dynamic range (DR) of the resulting images. Conversely, our imaging DR requirements effectively determine how much EGD we can tolerate in the system, and can thus be translated into beamformer calibration and/or hardware requirements. It is, however, very difficult to relate EGDs to the level of interferometric artefacts analytically. We have therefore decided to conduct a series of numerical simulations using the MeqTrees package [6]. The resulting simulations framework and methodology can be readily applied to other PAF designs, and even to aperture arrays.

\section{Simulations}

\paragraph*{Simulator inputs.} The inputs to our simulator are a set of element beam patterns, one per each element, 
and a collection of weight vectors $\vec w_{ij}$ (per each polarization direction $i=x,y$ and pointing $j=1\ldots37$) corresponding to the desired compound beams. These were computed using the system model and numerical tool in [2]. The other input is a sky model, which is essentially a list of sky sources to be simulated. For a given pointing, the simulator then works out the per-station compound beam gain in the direction of every source in the model, and evaluates a RIME that gives the resulting visibilities, which are written to a CASA-format Measurement Set (MS), which can then be imaged using standard tools (CASA or lwimager). Observational parameters, array layout, etc., are determined by the metadata of the MS. For this initial study, we simulated a 12-hour synthesis using the WSRT (with hypothetical APERTIF front-ends) at 1.42 GHz. Since we wanted to have a clear indication of how the relative effect varies as a function of direction, we used a ``gridded sky'' model of $11\times11$ identical unpolarized 1~Jy point sources, arranged in a grid with a step of $12'$ (thus extending out to $1^\circ$ from centre in both directions). Simulating real observations is just a matter of feeding the simulator a realistic sky model; for our purposes the gridded sky is actually more revealing.

\begin{figure}
\begin{tabular}{@{}cccc@{}}
\includegraphics[width=3.8cm]{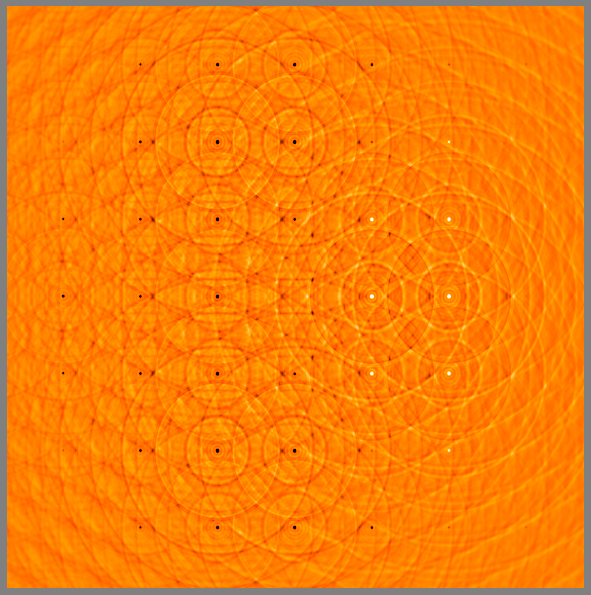} &
\includegraphics[width=3.8cm]{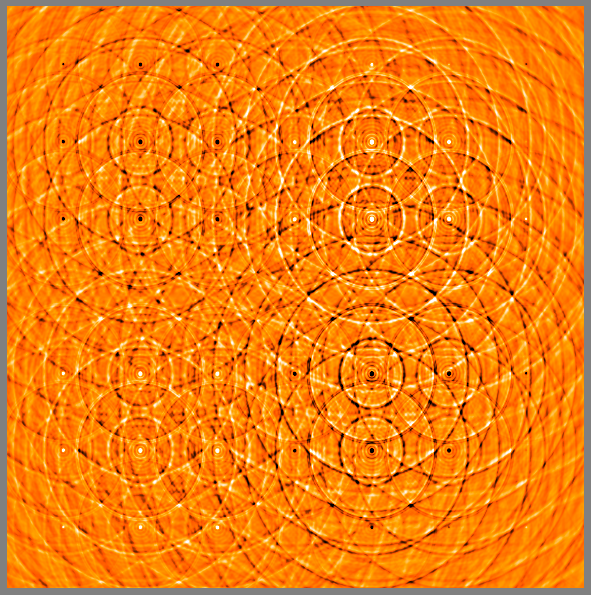} &
\includegraphics[width=3.8cm]{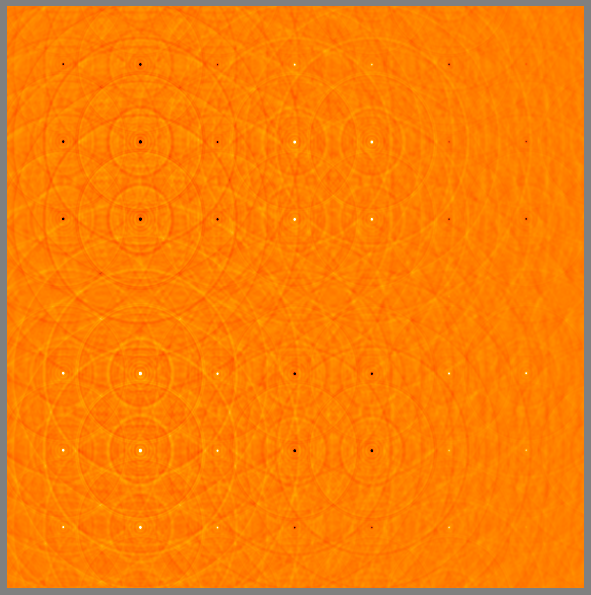} &
\includegraphics[width=3.8cm]{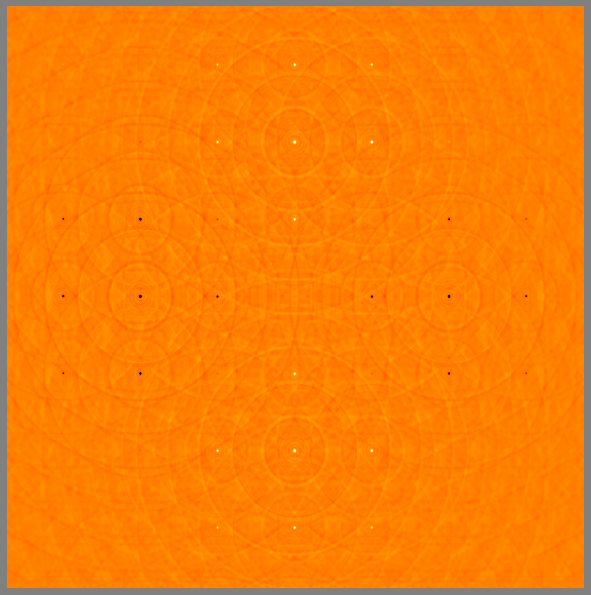} \\
$Q,$ beam \#16 & $U,$ beam \#16 & $V,$ beam \#16 & $Q,$ beam \#19
\end{tabular}
\caption{\label{fig:quv}$Q$, $U$, and $V$ maps showing a simulation without EGDs for beam \#16 (which is the leftmost off-axis beam in Fig.~\ref{fig:gains}a). Model sources are 1 Jy unpolarized, so this essentially shows the level of  instrumental polarization \emph{attenuated by the power beam}. The images are $1.5^\circ$ across. The intensity range shown is $\pm3$~mJy (actual extremes are on the order of 20~mJy). Note the relatively high $U$, and the asymmetry in all three images. This is characteristic of APERTIF's off-axis beams. For comparison, the rightmost pane shows a $Q$ map for the on-axis beam (\#19).}
\end{figure}

For the first (``reference'') simulation, we applied the nominal beam weights with no drift ($\mathbf{G}\equiv1$). The resulting $IQUV$ images are still quite informative, since they clearly show the distribution of instrumental polarization (due to different $x$ and $y$ beams, and cross-terms in the Jones matrix) over a single compound beam. Since the model sources have only $I$ flux, everything that makes its way into the $QUV$ maps is a result of instrumental polarization. Figure~\ref{fig:quv} shows the $QUV$ maps for beam \#16, which is one of the outermost beams, 3 HPBW off the optical axis of the dish. The most striking feature is how the polarimetric homogeneity of the beam patterns rapidly degrades as a function of direction. (Note that the images in effect show a combination of three things: fractional instrumental polarization, attenuation by the overall power beam, and convolution with the point spread function of the interferometer.) 

We then allowed $\mathbf{G}$ to drift as a function of time. For this initial study, we used a very primitive EGD model: the complex amplitude and phase of every element of $\mathbf{G}$ was modulated by a  sine wave, with a randomly-chosen period (between 2 and 6 hours) and initial phase. The amplitude was modulated by $\pm0.3$ dBA, and the phase by $\pm4$ degrees, which is consistent with real-life EGDs recently measured on the DIGESTIF prototype [3]. Each station's $\mathbf{G}$ was subject to its own modulation. A summary plot of the resulting compound beam gains (as a function of time) for a few directions and stations can be seen in Fig.~\ref{fig:gains}b. We call this the ``error'' simulation.

\begin{figure}
\begin{tabular}{@{}cccc@{}}
\includegraphics[width=3.8cm]{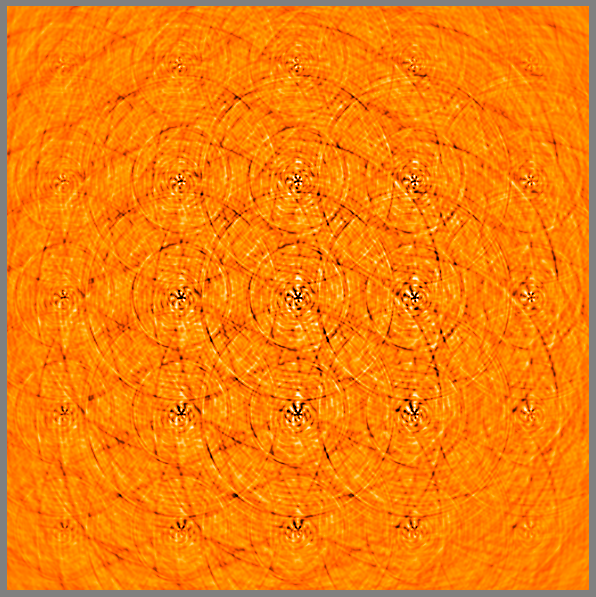} &
\includegraphics[width=3.8cm]{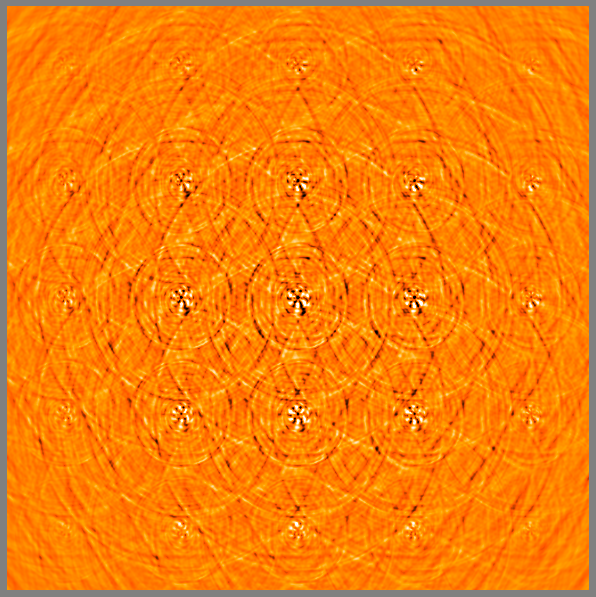} &
\hfill\includegraphics[width=3.8cm]{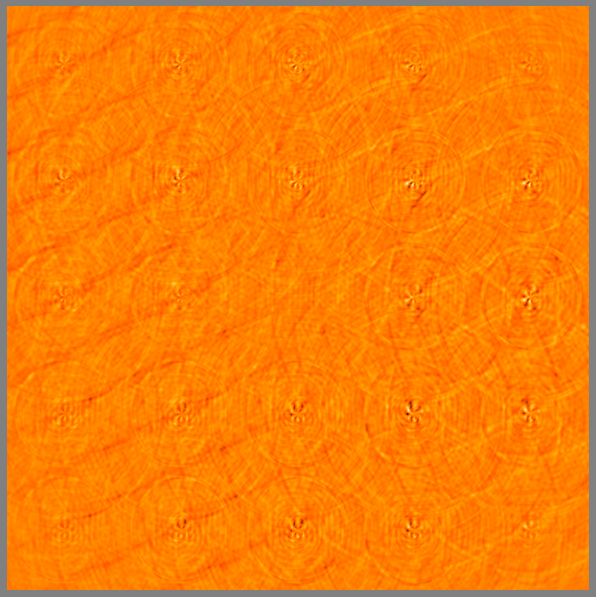} &
\hfill\includegraphics[width=3.8cm]{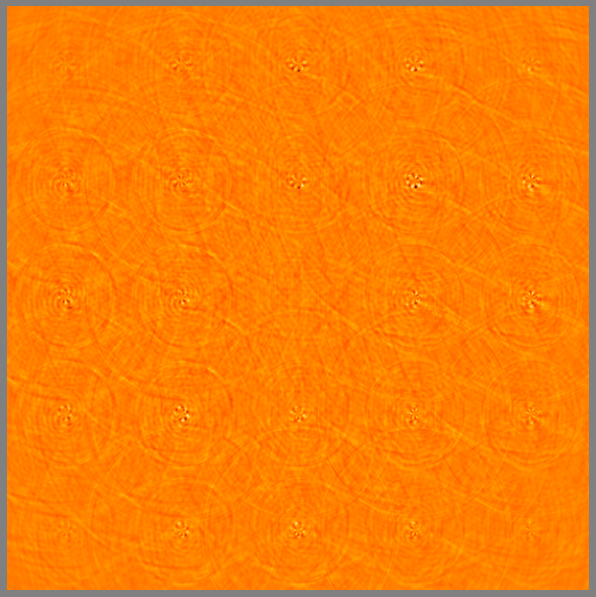} \\
(a) beam \#16&(b) beam \#19&(c) beam \#16&(d) beam \#19 \\
\multicolumn{2}{c}{(uncalibrated)} & \multicolumn{2}{c}{(post-selfcal)}
\end{tabular}
\caption{\label{fig:diff}(a,b) Imaging artefacts in Stokes $I$ due to EGDs. This shows the difference between the ``error'' and the ``reference'' images (see text) for beams \#16 and 19. Images (c) and (d) show the same differences after selfcal has been applied -- note how the central source is corrected for. The images are $1^\circ$ across. The intensity range shown is $\pm1$~mJy; actual extremes are on the order of 2~mJy in the uncalibrated images, and $\lesssim1$~mJy post-selfcal.}
\end{figure}

\paragraph*{Errors and dynamic range.} The resulting ``error'' images cannot be distinguished from the ``reference'' images by eye alone, since the effect is quite subtle. The difference between the two sets of images, however, is quite striking (Fig.~\ref{fig:diff}a,b). This indicates both the distribution of the artefacts, and their level, which corresponds to a DR limit of about 1:500. This is poor by radio interferometric standards, but can be improved by self-calibration. Given a dominant source, selfcal can estimate the overall gain \emph{in the direction of that source}, and correct for it. Any remaining artefacts are then due to the gain \emph{differences} between the dominant source and the other sources in the field, and increase (in relative level) as one gets further away from the dominant source. (These are often called ``off-axis'' artefacts, since conventional interferometric practice puts the dominant source in the centre of the beam, i.e. on-axis. For APERTIF this is not necessarily the case, so the more generic term ``DDEs'' is preferred.) 

We can simulate the effect of a ``perfect'' selfcal (for e.g. a dominant source at the centre) by dividing the beam gain in each direction by the beam gain towards the central source. This results in a set of ``corrected'' images. By imaging the difference between these and the ``reference'' images (Fig.~\ref{fig:diff}c,d), we get a clear indication of the artefact level after selfcal. Note how the artefacts go from none at centre (since that's the direction towards which the gain has been corrected), to a maximum somewhere around the half-power point, and then tail off again as the primary beam falls off. The corresponding DR limitation, post-selfcal, is on the order of 1:2000. This may still seem low, but recall that the dominant source would be taken care of by selfcal, so it is only the remaining sources that would leave artefacts on a \emph{relative} level of 1:2000. If these sources are relatively faint, the \emph{absolute} artefact level may be tolerably low. In fact, current single-pixel feed WSRT observations at 1.4 GHz exhibit an artefact level of 1:1000 on off-axis sources, which is thought to be due to pointing errors and other mechanical effects [7]. We can therefore conclude that, given the measured EGD levels, APERTIF's imaging DR will be limited by WSRT's existing optics and mechanics, rather than temporal instability of the primary beam pattern due to EGDs.

\section{Conclusions}

\noindent The tools and methodology presented here provides the missing link between the electromagnetic simulations of antenna engineers, and the  imaging requirements of astronomers. It is flexible enough that we can simulate wildly different PAF and AA configurations (including those with alt-az mounts), and determine how these ultimately influence dynamic range. In the particular case of APERTIF, our simulations together with measured EGD levels suggest that dynamic range will be limited by the WSRT hardware, and not the EGDs per se. 

\section{References}

\noindent 1. W. A. van Capellen and L. Bakker, ``APERTIF: Phased array feeds for the Westerbork Synthesis Radio Telescope'', \emph{Proceedings of the IEEE International Symposium on Phased Array Systems and Technology (ARRAY 2010)}, Waltham, MA, 2010, pp. 640--647

\noindent 2. M. V. Ivashina, O. A. Iupikov, R. Maaskant, W. A. van Capellen and T. Oosterloo, ``An Optimal Beamforming Strategy for Wide-Field Surveys With Phased-Array-Fed Reflector Antennas'', to appear in \emph{Special Issue on Antennas for Next Generation Radio Telescopes, IEEE Trans. on Antennas and Propagation}, 49, 2011, \#3

\noindent 3. W. A. van Cappellen and M. V. Ivashina, ``Temporal Beam Pattern Stability of Radio Astronomy Phased Array Feeds'', to appear in 
\emph{Proceedings of the 5th European Conference on Antennas and Propagation (EuCAP 2011)}, Rome, Italy, 2011

\noindent 4. J. P. Hamaker, ``Understanding radio polarimetry. IV. The full-coherency analogue of scalar self-calibration: Self-alignment, dynamic range and polarimetric fidelity''.
\emph{Astron. \& Astroph. Supp. Ser.}, 143, May 2000, pp. 515--534


\noindent 5. O. M. Smirnov, ``Revisiting the radio interferometer measurement equation. II. Calibration and direction-dependent effects'',
\emph{Astron. \& Astroph.}, 527, March 2011, p. A107

\noindent 6. J. E. Noordam and O. M. Smirnov, ``The MeqTrees software system and its use for third-generation calibration of radio interferometers'',
\emph{Astron. \& Astroph.}, 524, December 2010, p. A61

\noindent 7. O. M. Smirnov, ``Revisiting the radio interferometer measurement equation. III. Addressing direction-dependent effects in 21cm WSRT observations of 3C 147'',
\emph{Astron. \& Astroph.}, 527, March 2011, p. A108

\end{document}